\documentclass[11pt]{article}
\usepackage[a4paper]{geometry}
\usepackage{array}
\usepackage{amssymb}
\usepackage{amsfonts}
\usepackage{amsmath}
\usepackage{latexsym}
\usepackage{graphicx}
\usepackage{enumerate}
\usepackage{url}
\usepackage{nicefrac}
\usepackage{verbatim}
\usepackage[ruled,vlined]{algorithm2e}
\usepackage{mathtools}
\usepackage{xspace}

\def\mkvc{\textsc{max $k$-vertex cover}}
\def\mksc{\textsc{max $k$-set cover}\xspace}
\def\opt{\mathrm{opt}}

\def\OPT{\mathrm{OPT}}

\def\nphard{$\mathbf{NP}$-hard}

\newtheorem{THEOREM}{Theorem}
\newenvironment{theorem}{\begin{THEOREM} \hspace{-.85em} {\bf .} }%
                        {\end{THEOREM}}
\newtheorem{LEMMA}{Lemma}
\newenvironment{lemma}{\begin{LEMMA} \hspace{-.85em} {\bf .} }%
                      {\end{LEMMA}}
\newtheorem{COROLLARY}{Corollary}
\newenvironment{corollary}{\begin{COROLLARY} \hspace{-.85em} {\bf .} }%
                          {\end{COROLLARY}}
\newtheorem{PROPOSITION}{Proposition}
\newenvironment{proposition}{\begin{PROPOSITION} \hspace{-.85em} {\bf .} }%
                            {\end{PROPOSITION}}
\newtheorem{REMARK}{Remark}
\newenvironment{remark}{\begin{REMARK} \hspace{-.85em} {\bf .} \rm}%
                            {\end{REMARK}}

\newcommand{\thm}{\begin{theorem}}
\newcommand{\lem}{\begin{lemma}}
\newcommand{\pro}{\begin{proposition}}
\newcommand{\rem}{\begin{remark}}
\newcommand{\cor}{\begin{corollary}}
\newcommand{\prf}{\noindent{\bf Proof.} }
\newcommand{\ethm}{\end{theorem}}
\newcommand{\elem}{\end{lemma}}
\newcommand{\epro}{\end{proposition}}
\newcommand{\erem}{\bbox\end{remark}}
\newcommand{\ecor}{\end{corollary}}
\newcommand{\eprf}{\bbox}
\newcommand{\bbox}{\vrule height7pt width4pt depth1pt}

\let\leq\leqslant
\let\geq\geqslant

\begin{document}

\title{A polynomial time approximation schema for \mkvc{} in bipartite graphs}

\author{Vangelis Th. Paschos \\
Université Paris-Dauphine, PSL Research University \\
CNRS, UMR 7243 LAMSADE \\
\texttt{vangelis.paschos@lamsade.dauphine.fr}
}

\maketitle

\begin{abstract}
The paper presents a polynomial time approximation schema for the edge-weighted version of \mkvc{} problem in bipartite graphs. 
\end{abstract}

\section{Introduction}\label{intro}

The \mkvc{} problem is defined as follows: given a graph~$G(V,E)$ of order~$n$ and a constant $k \leqslant n$, determine~$k$ vertices that cover a maximum number of edges. In the (edge-)weighted version of the \mkvc{}, weights are associated with the edges and the objective is 
to determine~$k$ vertices that maximize the total weight of the edges covered by them. 
The problem is \nphard{} even in bipartite graphs~\cite{apollonio14,DBLP:journals/siamdm/CaskurluMPS17}.

\mkvc{} is a well-known restriction of \mksc{} problem where {we} are given a family of subsets~$\mathcal S$ over a set of elements~$C$ and an integer~$k$ and the objective is to determine a subfamily~$\mathcal S' \subseteq \mathcal S$ of cardinality~$k$, covering a maximum number of elements from~$C$. Analogously, in the weighted version of \mksc, the elements of~$C$ are provided with positive weights and the objective becomes to determine~$k$ sets that maximize the total weight of the covered elements.

Both \mksc{} and \mkvc{} are well-known  problems met in many real-world applications since they model several very natural facility location problems. In particular \mkvc{} is used for modeling problems in areas such as databases, social networks, sensor placement, information retrieval, etc. A non-exhaustive list of references to such applications can be found in~\cite{DBLP:conf/compgeom/BadanidiyuruKL12}.

To the best of our knowledge, the approximation of \mksc has been studied for the first time in the late seventies by Cornuejols et al.~\cite{cfn}, where an
approximation ratio $1-(\nicefrac{1}{e})$ ($\approx 0.632$) is proved for the natural greedy algorithm, consisting of iteratively choosing the currently largest-cardinality set, until~$k$ sets are included in the solution. \mkvc{} being a restriction of \mksc, the same ratio is achieved for the former problem also and this  ratio is tight in (weighted) bipartite graphs~\cite{DBLP:conf/compgeom/BadanidiyuruKL12}. A more systematic study of the greedy approximation of \mkvc{} can be found in~\cite{Hochbaum98}. In~\cite{Han02} it has been proved that the greedy algorithm also {achieves} ratio~$\nicefrac{k}{n}$. In the same paper, a very simple randomized algorithm is presented, that achieves approximation ratio $2(\nicefrac{k}{n}) - (\nicefrac{k}{n})^2$. Using a sophisticated linear programming method, the approximation ratio for \mkvc{}, in general graphs was improved to~$\nicefrac{3}{4}$~\cite{ageev}; this ratio remained the best known one in general graphs until~2018, when~\cite{DBLP:journals/corr/abs-1810-03792} proposed a 0.92-approximation for the problem. Obvously, this ratio remains valid for bipartite graphs. 
For this class, the best ratio (still based on linear programming) known before~\cite{DBLP:journals/corr/abs-1810-03792} was~$\nicefrac{8}{9}$~(\cite{DBLP:journals/siamdm/CaskurluMPS17}). The complexity of this~$\nicefrac{8}{9}$-approximation algorithm is not given in~\cite{DBLP:journals/siamdm/CaskurluMPS17}; a rough evaluation of it, gives a complexity of~$O((|L||R|)^{\nicefrac{11}{2}})$, where~$L$ and~$R$ are the two color-classes of~$B$ ($|L|+|R|=n$), which is bounded above by~$O(n^{11})$.
Finally, \mkvc{} being a generalisation of \textsc{min vertex cover}, its inapproximability by polynomial time approximation schemata (PTAS) in general graphs $\mathbf{P} = \mathbf{NP}$ is immediately derived from the corresponding inapproximability of the latter,~(\cite{DBLP:journals/cc/Patrank94}). 


Let us note that unweighted \mkvc{} is easy in semi-regular bipartite graphs (where all the vertices of each color class have the same degree). Indeed, any $k$ vertices in the color class of maximum degree yield an optimal solution. Obviously, if this color class contains less than $k$ vertices, then one can cover \emph{all} the edges.

In this paper, we propose a PTAS for \mkvc{} in bipartite graphs.

The following notations will be used in what follows:
\begin{itemize}
\item $B(L,R,E,\vec{w})$: an edge-weighted bipartite graph instance of \mkvc~($\vec{w}$ is a weight-vector of dimension~$|E|$);
\item $O$ an optimal solution of \mkvc;~$O_1$ and~$O_2$ the subsets of~$O$ lying in the color-classes~$L$ and~$R$, respectively;
\item $k_1 = |O_1|$ and~$k_2=|O_2|$ the numbers of the optimal vertices in~$L$ and~$R$, respectively; $k_1+k_2=k$; we assume that $k_1 \leqslant k_2$;
\item $\opt(B) = \opt$ the \textit{value} of an optimal solution (i.e., the value of the total coverage capacity of~$O$); $\opt(O_1)$ the \textit{total} coverage capacity of~$O_1$; we set $\opt(O_1) = \alpha\cdot\opt$, $\alpha \leqslant 1$;~$\opt(O_2)$ denotes the {\textit{private}} coverage capacity of~$O_2$, i.e., {\em the edges already covered by~$O_1$ are not encountered there}; obviously, $\opt(O_2) = (1-\alpha)\cdot\opt$;
\item for a vertex-set cardinality~$|X| \leqslant \opt$ ($X \subset O$):
\begin{itemize}
\item $C(X)\cdot\opt$ denotes the total weight of the edges covered by the members of~$X$ covering a maximum weight of edges 
(the best~$|X|$-elements subset of~$O$; $C(O_1)\cdot\opt= \alpha\cdot\opt$, $C(O_2)\cdot\opt = (1-\alpha)\cdot\opt$);
\item $C_w(X)\cdot\opt$ denotes the total weight of the edges covered by the members of~$X$ covering a minimum weight of edges (the worst~$|X|$-elements subset of~$O$); 
\end{itemize}
\item  $\rho$: the approximation ratio of a known \mkvc-algorithm.
\end{itemize}
The following proposition holds and will be frequently used in what follows.
\begin{proposition}\label{basic}
For any set~$X$ of optimal vertices: 
$$
\opt(O \setminus X) = \opt - C_w(X)\cdot\opt = \left(1 - C_w(X)\right)\cdot\opt
$$
\end{proposition}
Let us note that values of~$k_1$ and~$k_2$ can be guessed in polynomial time. We simply run the algorithm specified below for any possible pair of integers~$k'$ an~$k''$ such that $k'+k'' = k$ and take the best result of these runnings. One of the results will be obtained for the pair~$k_1,k_2$ and the solution returned will dominate the one for this particular pair. It is easy to see that this procedure takes, at worst,~$O(n^2)$ time. In what follows we will reason w.r.t. to the solution obtained for pair~$(k_1,k_2)$.

In what follows, we consider that the vertices of~$L$ and~$R$ are ordered in decreasing order w.r.t. their initial coverage capacity. Also, we call ``best'' vertices, a set of vertices that cover the largest total weight of \emph{uncovered} edges in~$B$.


\section{A preliminary result}

The following proposition shows that we can consider that the values of~$\vec{w}$ are polynomially bounded. Denote by \mkvc($n^{\ell}$) the instances of \mkvc{} where edge-weights are bounded above by~$n^{\ell}$, for a fixed constant ${\ell}\geqslant 3$.
\begin{proposition}\label{appresred}
There exists an approximation-preserving reduction between \mkvc{} and \mkvc($n^{\ell}$).
\end{proposition}
\prf
Consider an instance $B = (L,R,E,\vec{w})$ of \mkvc{} and produce an instance $B' = (L,R,E,\vec{w'})$ of \mkvc($n^{\ell}$) where~$L$, $R$ and~$E$ remain the same and any element~$w_i$ of~$\vec{w}$ is transformed in $w'_i = \lceil\nicefrac{n^{\ell}\cdot w_i}{w_{\max}}\rceil$ in~$B$, where~$w_{\max}$ is the maximum of the values in~$\vec{w}$. It is easy to see that all the elements in~$\vec{w'}$ are bounded above by~$n^{\ell}$ and that the transformation of~$B$ into~$B'$ can be done in polynomial time.

Assume now that there exists a polynomial $\rho$-approximation algorithm~\texttt{A} for \mkvc($n^{\ell}$) computing a solution~$S'$ consisting of~$m'$ edges of total weight~$\mathrm{sol}(B')$ and denote by~$\opt'$ the optimal value for the problem. Denote by~$\OPT$ the edge-set of an optimal solution of~$B$ and by $w^*_1, w^*_2, \ldots, w^*_{\OPT}$ the weights of its edges; obviously $\opt(B) = \sum_{i=1}^{\OPT}w^*_i$. We then have:
\begin{eqnarray}
\frac{n^{\ell}}{w_{\max}}\cdot\opt(B) &=& \frac{n^{\ell}}{w_{\max}}\cdot\sum_{i=1}^{\OPT}w^*_i \;\; \leqslant \;\; \sum_{i=1}^{\OPT}\left\lceil\frac{n^{\ell}\cdot w^*_i}{w_{\max}}\right\rceil \;\; = \;\;  \opt\left(B'\right)\label{opt'} \\
\mathrm{sol}\left(B'\right) &=& \sum_{i=1}^{m'}\left\lceil\frac{n^{\ell}\cdot w_i}{w_{\max}}\right\rceil \;\; \leqslant \;\; \frac{n^{\ell}}{w_{\max}} \cdot \sum_{i=1}^{m'}w_i + m' \nonumber \\
&\leqslant& \frac{n^{\ell}}{w_{\max}} \cdot \sum_{i=1}^{m'}w_i + \frac{n^2}{4} \;\; = \;\; \frac{n^{\ell}}{w_{\max}} \cdot \mathrm{sol}(B) + \frac{n^2}{4} \label{sol'}
\end{eqnarray}
where inequalities in~(\ref{sol'}) hold because the maximum number of edges in a bipartite graph of order~$n$ is bounded above by~$\nicefrac{n^2}{4}$.

Combining~(\ref{opt'}) and~(\ref{sol'}) and using the fact that \mkvc($n^{\ell}$) is polynomial time $\rho$-approximable, i.e., $\nicefrac{\mathrm{sol}(B')}{\opt(B')} \geqslant \rho$,  one can easily conclude that in this case \mkvc{} is polynomially approximable  within ratio $\rho - (\nicefrac{1}{(4\cdot n^{\ell-2})}) = \rho - \epsilon$.~\eprf


\section{Improving an approximation ratio~$\mathbf{\rho}$} 

The following proposition gives a lower bound for~$C_w(X)$, for any $X \subseteq L$, that will be used later (it is easy to see that, symmetrically, the same holds also for~$X \subseteq R$).

%
%
%
\begin{PROPOSITION}
\begin{equation}\label{cw}
C_w\left(X\right) \geqslant \frac{\rho - r + (1-\rho) \cdot C(X)}{\rho}
\end{equation}
where~$r$ is an upper bound of the approximation ratio of \texttt{Algorithm~\ref{alg1}}.
\end{PROPOSITION}
\prf
Denote by~$\texttt{A}$ a $\rho$-approximation algorithm for \mkvc, fix an integer $|X| \leq k_1$ and run \texttt{Algorithm~\ref{alg1}}. 

\begin{algorithm} 
\SetAlgoLined
\KwIn{A bipartite graph~$B(L,R,E)$ and a constant $k < |L|+|R|$}
\KwOut{A $k$-vertex cover of~$B$}
 {
    \begin{enumerate}
    \item\label{1.1} remove the best (first)~$|X|$ vertices from~$L$ together with their incident edges; let~$B'((L\setminus X),R,E',\vec{w})$ the remaining graph;
    \item\label{1.2} run~\texttt{A} on~$B'$ to solve \textsc{max ($k-|X|$)-vertex cover};
    \item\label{1.3} add to the result of step~\ref{1.2} the~$|X|$ vertices removed in step~\ref{1.1};
    \item return the so computed solution~SOL1.
    \end{enumerate}
    }
    \caption{}\label{alg1}
\end{algorithm}

Step~\ref{1.1} of Algorithm~\ref{alg1} deletes a subset~$X^*$ of optimal vertices members of~$O_1$ together with their (optimal) incident edges and, probably, another set of optimal edges incident to a subset of $\Gamma(X\setminus X^*)$ belonging to~$O^*_2$. Step~\ref{1.2} deletes from~$B$, $|X\setminus X^*|$ additional optimal vertices. After Steps~\ref{1.1} and~\ref{1.2}, the optimal value is at least:
$$
\left(1 - C(X) - C_w\left(X \setminus X^*\right)\right)\cdot\opt
$$
Finally, Step~\ref{1.3} produces a solution~SOL1 with value:
\begin{eqnarray}\label{ratio1f}
C\left(\mathrm{SOL}1\right) &\geqslant& \left[\rho\cdot\left(1 - C(X) - C_w\left(X \setminus X^*\right)\right)+C(X)\right]\cdot\opt \nonumber \\
&=& \left[\rho + (1-\rho)\cdot C(X) - \rho\cdot C_w\left(X \setminus X^*\right)\right]\cdot\opt
\end{eqnarray}
%
Since $C_w(X \setminus X^*) \leqslant C_w(X)$,~(\ref{ratio1f}) becomes:
$$
C\left(\mathrm{SOL1}\right) \geqslant \left[\rho + (1-\rho)\cdot C(X) - \rho\cdot C_w\left(X\right) \right]\cdot\opt
$$
Then:
$$
\left[\rho + (1-\rho) \cdot C(X) - \rho\cdot C_w\left(X\right) \right]\cdot\opt \leqslant r\cdot\opt \implies C_w\left(X\right) \geqslant \frac{\rho - r + (1-\rho) \cdot C(X)}{\rho}
$$
q.e.d.~\eprf

Consider now the following \texttt{Algorithm~\ref{alg2}}. 

\begin{algorithm} 
\SetAlgoLined
\KwIn{A bipartite graph~$B(L,R,E)$ and a constant $k < |L|+|R|$}
\KwOut{A $k$-vertex cover of~$B$}
 {
    \begin{enumerate}
    \item\label{2.1} fix a constant $c  > 2$;
    \item\label{2.2} \textbf{for} $\ell \leftarrow k_1$ \textbf{downto}~$c$ \textbf{do}
    \begin{enumerate}
    \item\label{2.2.1} run~\texttt{A} with $k := k' := k - \ell$; \\
    let~SOL($k'$) the solution obtained and~$m'_k$ its value;
    \item\label{2.2.3} complete SOL($k'$) by adding the~$\ell$ best vertices either from~$L$, or from~$R$ \\
    and store it;
    \end{enumerate}
    \item\label{2.3} \textbf{for} $\ell \leftarrow c$ \textbf{downto}~$1$ \textbf{do}
    \begin{enumerate}
    \item\label{2.3.1} compute all the sets on~$\ell$ vertices from~$B$ and store them together with \\
    the edges covered by any of them;
    \item\label{2.3.2} \textbf{for} any such set~$C$ \textbf{do} 
    \begin{enumerate}
    \item\label{2.3.2.1} delete it from~$B$ together with the edges covered by it and solve  \\
    \textsc{max $(k-\ell)$-vertex cover} in the remaining graph;
    \item\label{2.3.2.2} add~$C$ together with the edges covered by its vertices \\
    in the solution computed during the previous step, producing so \\
    a solution for \mkvc{} on~$B$; 
    \item\label{2.3.2.3} store the solution computed;
    \end{enumerate}
    \end{enumerate}
    \item\label{2.4} output the best among the solutions so computed.
    \end{enumerate}
    }
    \caption{}\label{alg2}
\end{algorithm}
%
\begin{THEOREM}
The approximation ratio of \texttt{Algorithm~\ref{alg2}} is:
$$
r \geqslant \frac{\rho + (1-\rho)^2}{1 + (1-\rho)^2}
$$
\end{THEOREM}
\prf
Consider Step~(\ref{2.2.3}) of \texttt{Algorithm~2}. If~SOL($k'$) leaves outside at least~$2\cdot \ell$ optimal vertices, then either~$L \setminus \mathrm{SOL}(k')$, or $R \setminus \mathrm{SOL}(k')$ contains a set~$\mathcal{L}^{\ell}$ of at least~$\ell$ optimal vertices with a coverage capacity at least~$C_w(\mathcal{L}^{\ell})$. So, in this case, by adding the~$\ell$ best vertices either from~$L$, or from~$R$ and taking into account Proposition~\ref{basic}, \texttt{Algorithm~2} builds a solution with value at least:
\begin{eqnarray}\label{case1}
m'_k + C_w\left(\mathcal{L}^{\ell}\right)\cdot\opt &\geqslant& \rho\cdot\left(1 - C_w\left(\mathcal{L}^{\ell}\right)\right)\cdot\opt + C_w\left(\mathcal{L}^{\ell}\right)\cdot\opt \nonumber \\
&=& \left(\rho + (1-\rho)\cdot C_w\left(\mathcal{L}^{\ell}\right)\right)\cdot\opt 
\end{eqnarray}
Let now assume that Step~\ref{2.2.3} of \texttt{Algorithm~2} leaves outside less than~$2\cdot \ell$ optimal vertices.
If there exist~$\ell$ vertices (outside of the solution computed) either in~$L$, or in~$R$, with coverage capacity at least~$C_w(\mathcal{L}^{\ell})$, then the solution computed in Step~\ref{2.2.3} adds in the solution an additional coverage capacity at least~$C_w(\mathcal{L}^{\ell})$ and the discussion just above always holds.

So, the case remaining to be handled, is the one where:
$$
m'_k \geqslant \left[1 - 2\cdot C_w\left(\mathcal{L}^{\ell}\right)\right]\cdot\opt
$$
Then:
\begin{equation}\label{case2}
\left[1 - 2\cdot C_w\left(\mathcal{L}^{\ell}\right)\right]\cdot\opt \leqslant m'_k \leqslant r\cdot \opt \implies C_w\left(\mathcal{L}^{\ell}\right)  \geqslant \frac{1-r}{2}
\end{equation}
For example, suppose that~(\ref{case1}) occurs for $\ell = k_1$ ($k' = k_2$). Then, following~(\ref{cw}):
\begin{equation}\label{case1a}
C_w(\mathcal{L}) \geqslant \frac{\rho -r + (1-\rho)\cdot\alpha}{\rho}
\end{equation}
and embedding~(\ref{case1a}) in~(\ref{case1}), one obtains a ratio:
$$
r \geqslant \rho + \frac{\rho -r + (1-\rho)\cdot\alpha}{\rho}\cdot(1-\rho)
$$
which, after some easy algebra becomes:
\begin{equation}\label{ratiok1}
r \geqslant \frac{\rho - r \cdot (1-\rho) + (1-\rho)^2\cdot \alpha}{\rho}
\end{equation}
Consider now the very simple algorithm consisting of taking the~$k$ first best vertices of~$R$ (recall that vertices in ~$R$ are ordered in decreasing order with respect to their coverage capacity). It guarantees at least $1-\alpha$.  So:
\begin{equation}\label{ratio1-a}
1 - \alpha \leqslant r \implies \alpha \geqslant 1 - r
\end{equation}
Combining~(\ref{ratiok1}) and~(\ref{ratio1-a}) one gets, after some easy algebra:
\begin{equation}\label{ratio1}
r \geqslant \frac{\rho + (1-\rho)^2}{1 + (1-\rho)^2} > \rho
\end{equation}
Let now consider the other extreme case where Step~(\ref{2.2.3}) of \texttt{Algorithm~2} leaves outside less than~$2\cdot \ell$ optimal vertices for any~$\ell$ from~$k_1$ down to~$c$. 
In this case~(\ref{case2}) holds for any execution of Step~\ref{2.2} of the algorithm. Then, in Step~\ref{2.3}, \texttt{Algorithm~2} guarantees ratio:
$$
r \geqslant \rho \cdot \left(1 - C_w\left(\mathcal{L}^{\ell}\right)\right)+  C_w\left(\mathcal{L}^{\ell}\right) = \rho + (1-\rho)C_w\left(\mathcal{L}^{\ell}\right)  \overset{(\ref{case2})}{\geqslant} \rho + (1-\rho)\cdot\frac{1-r}{2}
$$
which gives:
\begin{equation}\label{ratio2}
r \geqslant \frac{1+\rho}{3-\rho} \geqslant \rho
\end{equation}
Assume now that from $\ell = k_1$ to $\ell = k_1'$, Step~(\ref{2.2.3}) of \texttt{Algorithm~2} leaves outside less than~$2\cdot \ell$ optimal vertices and that for $\ell = k_1' - 1$ Step~(\ref{2.2.3}) of \texttt{Algorithm~2} leaves outside more than~$2\cdot \ell$ optimal vertices.  This means (following~(\ref{case2})) that until the $k'_1$-th execution of Step~\ref{2.2}, $C_w^{k'_1}(\mathcal{L}^{k'_1})  \geqslant \nicefrac{(1-r)}{2}$ and at the $(k'_1+1)$-th execution ($\ell = k_1' - 1$),~$C_w^{k'_1+1}(\mathcal{L}^{k'_1+1})$ becomes smaller than~$\nicefrac{(1-r)}{2}$. 
Denote by~$w(v)$ the coverage capacity of a vertex $v \in V$ and consider the sum $C_w^{k'_1+1}(\mathcal{L}^{k'_1+1}) + w(v)$, for any $v \in V \setminus \mathcal{L}^{k'_1}$. Obviously, there exists at least a $v \in \mathcal{L}^{k'_1} \setminus \mathcal{L}^{k'_1+1}$  such that: $C_w^{k'_1+1}(\mathcal{L}^{k'_1+1}) + w(v) \geqslant C_w^{k'_1}(\mathcal{L}^{k'_1}) $, which implies $C_w^{k'_1+1}(\mathcal{L}^{k'_1+1}) \geqslant C_w^{k'_1}(\mathcal{L}^{k'_1}) -  w(v)$. We can assume that $w(v) \leq (\nicefrac{(1-r)}{4})\cdot\opt$. Indeed, if $w(v) \geq  (\nicefrac{(1-r)}{4})\cdot\opt$, then Step~\ref{2.3} of \texttt{Algorithm~2} computes a set~$C$ of coverage capacity at least~$(\nicefrac{(1-r)}{4})\cdot\opt$ and then discussion above concludes to a ratio:
\begin{equation}\label{ratio3}
r \geqslant \rho + (1-\rho)\cdot\frac{1-r}{4} \implies r \geqslant \frac{1+3\cdot\rho}{5 - \rho} > \rho
\end{equation}
Assuming $w(v) \leq (\nicefrac{(1-r)}{4})\cdot\opt$ we get:
\begin{equation}\label{transition}
C_w^{k'_1+1}\left(\mathcal{L}^{k'_1+1}\right)  \geqslant C_w^{k'_1}\left(\mathcal{L}^{k'_1}\right) - \frac{(1-r)}{4}
\overset{(\ref{case2})}{\geqslant} \frac{1-r}{4}
\end{equation}
Following the discussion above, for the value~$m$ of the so-produced solution holds:
\begin{eqnarray*} 
m &\geqslant& \left[\rho\cdot\left(1 - C_w^{k'_1+1}\left(\mathcal{L}^{k'_1+1}\right)\right)+C_w^{k'_1+1}\left(\mathcal{L}^{k'_1+1}\right)\right]\cdot\opt
\nonumber \\
&\overset{(\ref{transition})}{\geqslant} & \left(\rho + (1-\rho)\cdot\frac{1-r}{4}\right)\cdot\opt
\end{eqnarray*}
Solving the inequality:
$$
\left(\rho + (1-\rho)\cdot\frac{1-r}{4}\right) \leqslant r
$$
one gets:
\begin{equation}\label{ratio4}
r \geqslant \frac{1+3\cdot\rho}{5 - \rho} > \rho
\end{equation}
Some very simple algebra concludes that the lower bound for~$r$ given in~(\ref{ratio1}) is the smallest among the ones given in~(\ref{ratio1}), (\ref{ratio2}), (\ref{ratio3}) and~(\ref{ratio4}).

The proof of the theorem is now concluded.~\eprf

\section{A PTAS for \mkvc}


\begin{algorithm} 
\SetAlgoLined
\KwIn{A bipartite graph~$B(L,R,E)$ and a constant $k < |L|+|R|$}
\KwOut{A $k$-vertex cover of~$B$}
 {
    fix a constant  $\epsilon > 0$\;
    \For{i:= 1 \textbf{to} $(\nicefrac{2\cdot\epsilon\cdot(1-\epsilon-\rho)}{(1 - 2\cdot\epsilon^2-\sqrt{(1-4\cdot\epsilon^2)})})$}{
    run \texttt{Algorithm~\ref{alg2}} to solve \mkvc{} on~$B$ by using \texttt{Algorithm~\ref{alg2}}  itself as~\texttt{A}\;
    *for $i=1$~\texttt{A} is used* \\
    }
    \Return the solution computed.
    }
    \caption{A polynomial time approximation schema for \mkvc{} in bipartite graphs}\label{ptas}
\end{algorithm}

Revisit~(\ref{ratio1}). Some easy algebra allows to conclude that:
\begin{equation}\label{r-rho}
r - \rho \geqslant \frac{(1 - \rho)^3}{1 + (1-\rho)^2}
\end{equation}
and this quantity is decreasing with~$\rho$.

Consider now \texttt{Algorithm~\ref{ptas}} and denote by~$\rho_i$, the ratio of~\texttt{A} ($\rho_i = \rho$), by~$\rho_j$, the ratio of Algorithm~\ref{ptas} after the $j$-th execution of the loop \textbf{for} and by~$\rho_f$ the ratio of~Algorithm~\ref{ptas} after the last execution.  Suppose also that the \textbf{for} loop of Algorithm~\ref{ptas} is executed~$t$ times. Then, taking~\ref{r-rho}, into account:
\begin{equation}\label{rf-r}
\rho_f - \rho \geqslant t\times \left(\rho_f - \rho_{f-1}\right)
\end{equation}
For facility, in what follows, we work with equality in~(\ref{ratio1}), we set $\rho_f = 1 - \epsilon$ and suppose that the loop \textbf{for} of Algorithm~\ref{ptas} is executed~$t$ times. Setting in~(\ref{ratio1}), $r = \rho_f = 1 - \epsilon$ and $\rho = \rho_{f-1}$ we have, after some easy algebra:
\begin{eqnarray}\label{rf-1}
1 - \epsilon = \frac{\rho_{f-1} + \left(1-\rho_{f-1}\right)^2}{1 + \left(1-\rho_{f-1}\right)^2} &\implies& \rho_{f-1} = \frac{-(1 - 2\cdot\epsilon) + \sqrt{1 - 4\cdot\epsilon^2}}{2\cdot\epsilon} \nonumber  \\
&\implies& \rho_f - \rho_{f-1} = \frac{1 - 2\cdot{\epsilon^2 - \sqrt{1 - 4\cdot\epsilon^2}}}{2 \cdot\epsilon}
\end{eqnarray}
Combining~(\ref{rf-r}) and~(\ref{rf-1})
$$
(1-\epsilon) - \rho \geqslant t\cdot \left(\frac{1 - 2\cdot\epsilon^2 - \sqrt{1 - 4\cdot\epsilon^2}}{2 \cdot\epsilon}\right) \implies t \leqslant \frac{2\cdot\epsilon\cdot(1 - \epsilon - \rho)}{1 - 2\cdot\epsilon^2 - \sqrt{1 - 4\cdot\epsilon^2}}
$$
Since the discussion just above holds for any fixed constant $\epsilon < 1-\rho$, the following theorem holds immediately.
\begin{theorem}
\mkvc{} in bipartite graphs admits a polynomial time approximation schema.
\end{theorem}

\paragraph{Acknowledgement.}  The very useful discussions with Cécile Murat, Federico Della Croce, Michael Lampis and Aristotelis Giannakos are gratefully acknowledged.


\begin{thebibliography}{1}

\bibitem{ageev}
A.~A. Ageev and M.~Sviridenko.
\newblock Approximation algorithms for maximum coverage and max cut with given
  sizes of parts.
\newblock In G.~Cornu{\'e}jols, R.~E. Burkard, and G.~J. Woeginger, editors,
  {\em Proc. Conference on Integer Programming and Combinatorial Optimization,
  IPCO'99}, volume 1610 of {\em Lecture Notes in Computer Science}, pages
  17--30. Springer-Verlag, 1999.

\bibitem{apollonio14}
N.~Apollonio and B.~Simeone.
\newblock The maximum vertex coverage problem on bipartite graphs.
\newblock {\em Discrete Appl. Math.}, 165:37--48, 2014.

\bibitem{DBLP:conf/compgeom/BadanidiyuruKL12}
A.~Badanidiyuru, R.~Kleinberg, and H.~Lee.
\newblock Approximating low-dimensional coverage problems.
\newblock In T.~K. Dey and S.~Whitesides, editors, {\em Proc. Symposuim on
  Computational Geometry, SoCG'12, Chapel Hill, NC}, pages 161--170. {ACM},
  2012.

\bibitem{DBLP:journals/siamdm/CaskurluMPS17}
B.~Caskurlu, V.~Mkrtchyan, O.~Parekh, and K.~Subramani.
\newblock Partial vertex cover and budgeted maximum coverage in bipartite
  graphs.
\newblock {\em {SIAM} J. Discrete Math.}, 31(3):2172--2184, 2017.

\bibitem{cfn}
G.~Cornuejols, M.~{L.} Fisher, and G.~{L.} Nemhauser.
\newblock Location of bank accounts to optimize float: an analytic study of
  exact and approximate algorithms.
\newblock {\em Management Sci.}, 23(8):789--810, 1977.

\bibitem{Han02}
Q.~Han, Y.~Ye, H.~Zhang, and J.~Zhang.
\newblock On approximation of max-vertex-cover.
\newblock {\em European J.~Oper. Res.}, 143:342--355, 2002.

\bibitem{Hochbaum98}
D.~S. Hochbaum and A.~Pathria.
\newblock Analysis of the greedy approach in problems of maximum $k$-coverage.
\newblock {\em Naval Research Logistics}, 45:615--627, 1998.

\bibitem{DBLP:journals/corr/abs-1810-03792}
Pasin Manurangsi.
\newblock A note on max k-vertex cover: faster fpt-as, smaller approximate
  kernel and improved approximation.
\newblock {\em CoRR}, abs/1810.03792, 2018.

\bibitem{DBLP:journals/cc/Patrank94}
E.~Petrank.
\newblock The hardness of approximation: gap location.
\newblock {\em Computational Complexity}, 4:133--157, 1994.

\end{thebibliography}

\end{document}